# Fast WDM provisioning with minimal probing: the first field experiments for DC exchanges

HIDEKI NISHIZAWA,[1,*] TORU MANO,[1] THOMAS FERREIRA DE LIMA,[2] YUE-KAI HUANG,[2] ZEHAO WANG,[3] WATARU ISHIDA,[4] MASAHISA KAWASHIMA,[5] EZRA IP,[2] ANDREA D'AMICO,[6] SEIJI OKAMOTO,[1] TAKERU INOUE,[1] KAZUYA ANAZAWA,[1] VITTORIO CURRI,[6] GIL ZUSSMAN,[7] DANIEL KILPER,[8] TINGJUN CHEN,[3] TING WANG,[2] KOJI ASAHI,[9] KOICHI TAKASUGI[1]

*[1]NTT Network Innovation Labs, Kanagawa, Japan, [2]NEC Labs America, Princeton, USA, [3]Duke University, Durham, NC, USA, [4]NTT Software Innovation Center, Tokyo, Japan, [5]NTT IOWN Development Center, Tokyo, Japan, [6]Politecnico di Torino, Torino, Italy,[76]Columbia University, New York, USA,[8]CONNECT Centre, Trinity College Dublin, Ireland, [9]NEC Corporation, Chiba, Japan.*
**Corresponding author: hideki.nishizawa@ntt.com*

**There are increasing requirements for data center interconnection (DCI) services, which use fiber to connect any DC distributed in a metro area and quickly establish high-capacity optical paths between cloud services and mobile edge computing and the users. In such networks, coherent transceivers with various optical frequency ranges, modulators, and modulation formats installed at each connection point must be used to meet service requirements such as fast-varying traffic requests between user computing resources. This requires technology and architectures that enable users and DCI operators to cooperate to achieve fast provisioning of WDM links and flexible route switching in a short time, independent of the transceiver's implementation and characteristics. We propose an approach to estimate the end-to-end (EtE) GSNR accurately in a short time, not by measuring the GSNR at the operational route and wavelength for the EtE optical path but by simply applying a quality of transmission probe channel link by link, at a wavelength/modulation-format convenient for measurement. Assuming connections between transceivers of various frequency ranges, modulators, and modulation formats, we propose a device software architecture in which the DCI operator optimizes the transmission mode between user transceivers with high accuracy using only common parameters such as bit error rate (BER). In this paper, we first implement software libraries for fast WDM provisioning and experimentally build different routes to verify the accuracy of this approach. For the operational EtE GSNR measurements, the accuracy estimated from the sum of the measurements for each link was 0.6 dB, and the wavelength-dependent error was about 0.2 dB. Then, using field fibers deployed in the NSF COSMOS testbed, a Linux-based transmission device software architecture, and transceivers with different optical frequency ranges, modulators, and modulation formats, the fast WDM provisioning of an optical path was completed within 6 minutes.**

## 1. Introduction

### A. Emerging Expectations for DCX

As cloud services are expanding, there have been increasing efforts to relocate data centers previously concentrated in urban areas to suburban areas with abundant power and infrastructure space to equalize regional energy supplies. Sovereign clouds, which keep data within a designated geographic area to improve data security and privacy, are also becoming more prevalent. In addition, the growing use of private 5G technology has increased the need for high-speed, low-latency optical links between remote locations. In paticular, there are increasing cases where data owned in local data centers are connected to cloud services and mobile edge computing (MEC) using dark fiber and high-capacity optical links. However, these demands are met with significant challenges, such as the time it takes to procure dark fiber and start using the network, and the complexity and cost of the network increases as the number of locations to be connected increases.

Against this backdrop, there are growing expectations for data center exchange (DCX) services that directly connect any DCs distributed in a metro area via optical fiber to quickly establish high-capacity optical paths between each company's cloud services and MECs and the users connected to them [1].

Fig. 1 shows an example of a DCX configuration. Urban data centers (UDCs) are located in the gray area in the center of Fig. 1, and they are connected by carrier links (CLs) owned by the carrier, which provides a point of presence (POP) in each UDC.

User equipment can connect from a suburban data center (SDC) to the POP in UDC via alien access links (AALs) and directly to a remote cloud MEC or other user equipment without electrical conversion. Here, the user equipment can be located at either an SDC or UDC, and if it is at a UDC, it connects directly to the POP without an access line.

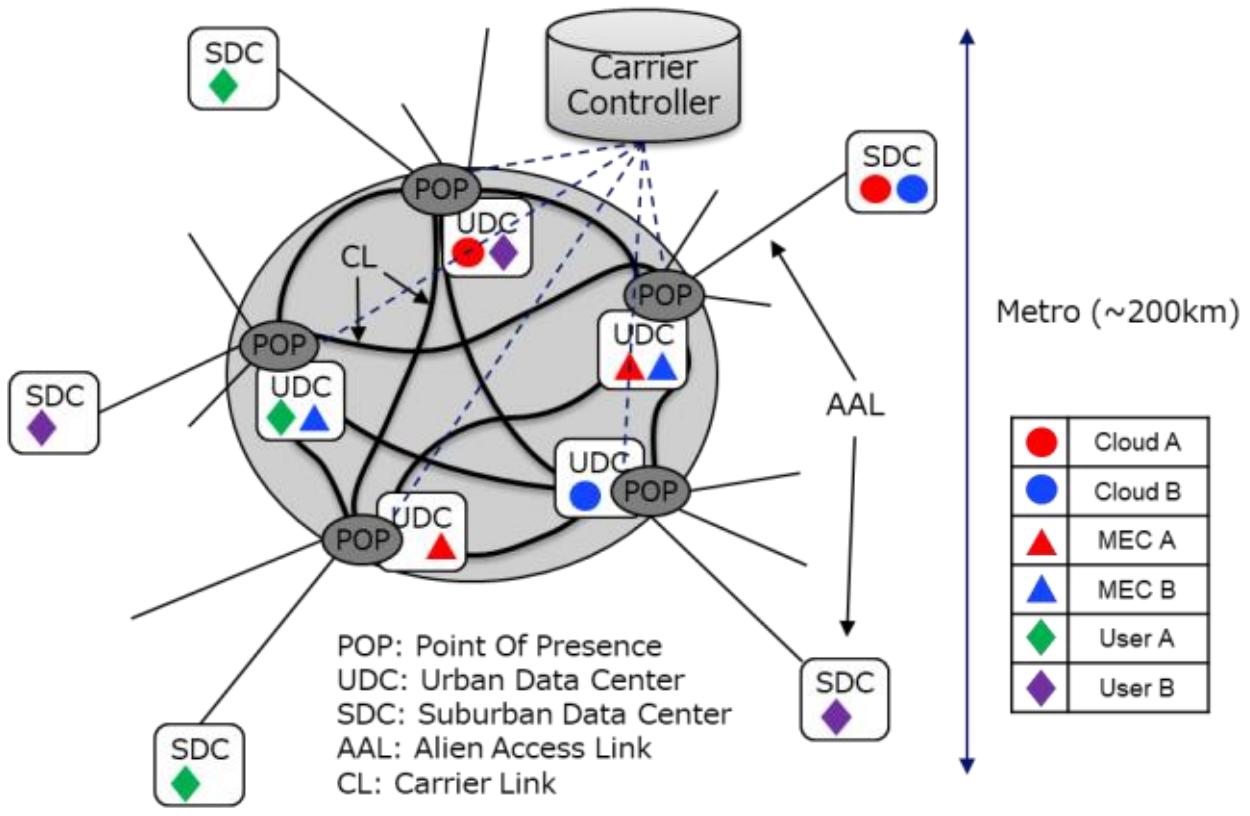

Fig. 1. Example of DCX physical deployment.

### B. Service Requirements for DCX

Several challenges still remain in the development of the DCX services that require real-time WDM provisioning to quickly fulfill user demand, and protocols to enable user-carrier cooperation automatically. Also, when a failure occurs in a user TRx installed outside the carrier network, the carrier needs to isolate the root cause and perform restoration work. The service requirements for DCX are broken down as follows:

(1) Rapid response to fast time-varying traffic requests that occur between DCs;
(2) Prompt optical path restoration in the event of disasters such as earthquakes and floods;
(3) Support for ultra-low latency services for inter-computer communications;
(4) Direct connection between various devices installed at user sites, cloud, and MECs without optical-electrical conversion; and
(5) Monitoring and control of user equipment by carriers to ensure stable operation of network services.

Here, the target response time required for (1) and (2) is set to be around 10 minutes (For (2), the case where the quality of transmission (QoT) of multiple optical links are degraded at once is included, this target is defined as the time required to restore one end-to-end (EtE) optical path between a particular user's muxponder). The latency in (3) is assumed to be within 2 ms, dominated by fiber propagation delay. This delay corresponds to a metro area of roughly 200 km in radius, depending on the fiber installation conditions.

### C. Importance of Open Interface/Specification

An open architecture that enables fast WDM provisioning between various devices must be designed to address requirements (1)-(4) described in Section B. The architecture should allow DCX operators to exchange information to calculate the QoT of each link in real-time, independent of the vendor or implementation of each device. This requires accurately estimating the generalized signal-to-noise ratio (GSNR) that indicates the signal quality at the physical layer in all possible transmission channels between the user sites.

For requirements (4) and (5) described in Section B, interoperability is a crucial among users, clouds, and MECs because each player uses their preferred devices which are adapted to their system and operation.

Standardization organizations such as the Optical Internetworking Forum (OIF) [2], Open ROADM MSA [3], and OpenZR+ MSA [4] have defined common hardware specifications for data plane interoperability. For control plane interoperability, there is no common specification that addresses the service requirements for DCX within the optical transport industry, while OpenROADM MSA, Telecom Infra Project Open Optical & Packet Transport (TIP OOPT) [5, 6, 7], and IOWN Global Forum (IGF) [8] provide architectures and open interfaces that partially meet these needs identified here.

### D. Fast WDM Provisioning Approach based on Open Architecture

In this paper, we propose a fast WDM provisioning approach for WDM links, including two types of commercial TRxs implemented with indium phosphide (InP) and silicon photonics (SiP) modulators. Our approach uses a technique to estimate the GSNR of all possible EtE optical paths quickly and accurately by only using a probe WDM channel that measures QoT link by link at measurable, convenient arbitrary modulation formats and modulators.

We have verified a minimal probing method and an integrated Linux-based device software architecture that utilizes open interfaces, specifications, and architectures defined by Open ROADM MSA, TIP OOPT, and IGF. We leverage Open ROADM MSA-compliant coherent TRxs to ensure data plane interconnectivity. For the hardware abstraction interface and network operating system (NOS) that controls user TRxs, we applied the TAI architecture and the Goldstone NOS that are under development openly in TIP OOPT. The architecture proposed in the IGF was adopted for the network architecture of user TRx and POP. We implement software libraries for fast WDM provisioning, and monitoring/configuration of the user's TRx to provide control plane interoperability. To estimate EtE GSNR accurately and design the optimal optical path in a short time, it was necessary to automatically obtain the back-to-back TRx characteristic because this is the most critical parameter for improving accuracy in metro areas. The proposed device software architecture utilizes server-based technologies such as containers and a hardware abstraction interface for coherent modules, and then solves the issue of automatic obtaining of the back-to-back TRx characteristic currently not supported by standard YANG models, such as the Open ROADM YANG model and the OpenConfig YANG model.

In addition, we experimentally verified the accuracy of this approach using three different routes of 32 km/72 km/122 km in the C-band. For the operational EtE GSNR measurements, the accuracy estimated from the sum of the measurements for each link was 0.6 dB, and the wavelength-dependent error was about 0.2 dB. Then, using field fibers deployed in an urban area, a Linux-based transmission device software architecture, and coherent transceivers with different optical frequency ranges, modulators, and modulation formats, the fast WDM provisioning of an optical path was completed within 6 minutes.

The remainder of this paper is organized as follows. After introducing related work in Section 2, we introduce our approach to estimating the GSNR of all possible EtE optical paths quickly with high accuracy in Section 3. Section 4 presents a Linux-based muxponder device software architecture that includes a DCX operator agent as a container. The architecture also allows for optimization of different settings of coherent transceivers (TRxs) from multiple vendors with different optical frequency ranges and modulators. In Section 5, we evaluate and compare the estimation error of this method and the penalty due to wavelength dependence and wavelength assignment of the device against the actual EtE GSNR measurements. In Section 6, we report the time measurements for the quality estimation of each link, EtE optical path margin design, and remote configuration of user muxponders using an experimental setup containing field fibers. Section 7 discusses future challenges for improving accuracy and expanding system flexibility, and Section 8 concludes the paper.

## 2. Related work

A converged inter/intra data center network architecture with an autonomous control plane for flexible bandwidth allocation has been introduced as a way to quickly establish high-capacity optical paths between distributed DCs in a metro area [9]. The architecture includes two types of physical layer connections (Background and Dynamic) to meet stringent bandwidth and delay requirements. Subsequently, Ju et al. targeted the provisioning of disaster-resilient cloud services using distance adaptive modulation formats by simultaneously considering content placement, modulation-format adaptive routing, path and content protection, and spectrum allocation [10]. They investigated frequency utilization with different traffic parameters and different network topologies by numerical simulations. However, neither of these prior studies considered WDM provisioning.

Kaeval et al. has conducted a number of studies on transmission mode optimization approaches for WDM provisioning [11,12,13]. They proposed an approach for provisioning WDM using QoT probe channels in a transparent optical path connecting two endpoints in a single or multi-domain optical network, referred to as optical spectrum as a service (OSaaS). Kaeval et al. estimated the GSNR in EtE between the TRxs of OSaaS users [11,12] and proposed an approach to estimate the GSNR EtE using the estimated GSNR for each link that constitutes the EtE route [13]. In both cases, the GSNR is estimated for each wavelength using a probe channel with the same wavelength as the signal light in operation, which requires time for provisioning in the case of multi-wavelength multiplexed transmission. Furthermore, while OSaaS is a P-to-P connection, DCX is a many-to-many connection deployed on a two-dimensional plane, as shown in Fig. 2(b), so the model is more complex than that in [13], with many combinations of optical path routes and wavelengths. The differences between the models will be discussed in detail in Section 3. OSaaS does not provide an architecture for users and carriers to work together when setting WDM paths. Furthermore, a single vendor device is applied to the transmitter and receiver, and no study or experiment has been conducted to set up optical paths between optical transmission devices with various optical frequency ranges, modulators, and modulation formats installed at each connection point, such as DCX. This will be discussed in detail in Section 4.

An architecture for users and carriers to work together using optical transmission devices with optical frequency range and modulation format has been proposed in [14]. In the proposed architecture and protocol, a mode catalog, including the modulation format and characteristics of the coherent TRx, is generated at the user site, the QoT of the link is estimated, and these are sent to the carrier's controller to optimize and configure the user site equipment. In [15], it was determined that the back-to-back characteristics of coherent TRxs can be accurately modeled for different types of coherent implementations and modulators. However, [14] and [15] do not mention network topology, specific steps for mode optimization, or a device software architecture for user TRx remote control, and only laboratory experiments have been conducted.

## 3. Fast and accurate GSNR estimation of all possible EtE optical paths

This section presents an approach to WDM provisioning and selecting the optimal mode in a short time, which is essential for satisfying service requirements (1)–(4).

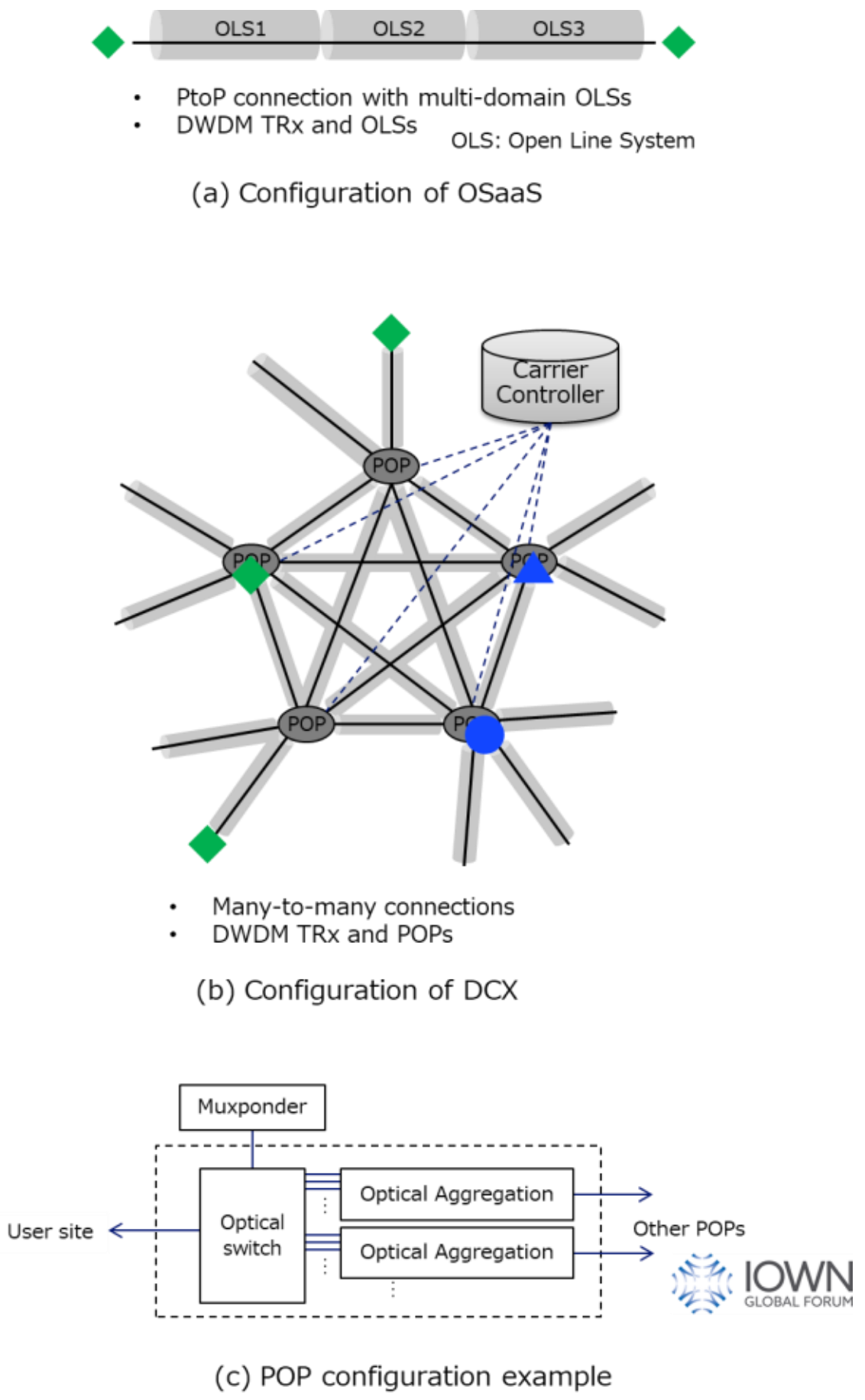

Fig. 2. Example network configuration of DCX.

Fig. 2(a) shows the OSaaS configuration [11], which consists of a multi-domain open line system (OLS) and can accommodate many users by applying DWDM. Fig. 2(b) shows an example of a network forming a DCX, in which a functionally disaggregated ROADM device [8] consisting of an optical switch and optical aggregation is applied to the POP in Fig. 2(c). As in Fig. 2(a), DCX can efficiently accommodate many user groups with DWDM technologies.

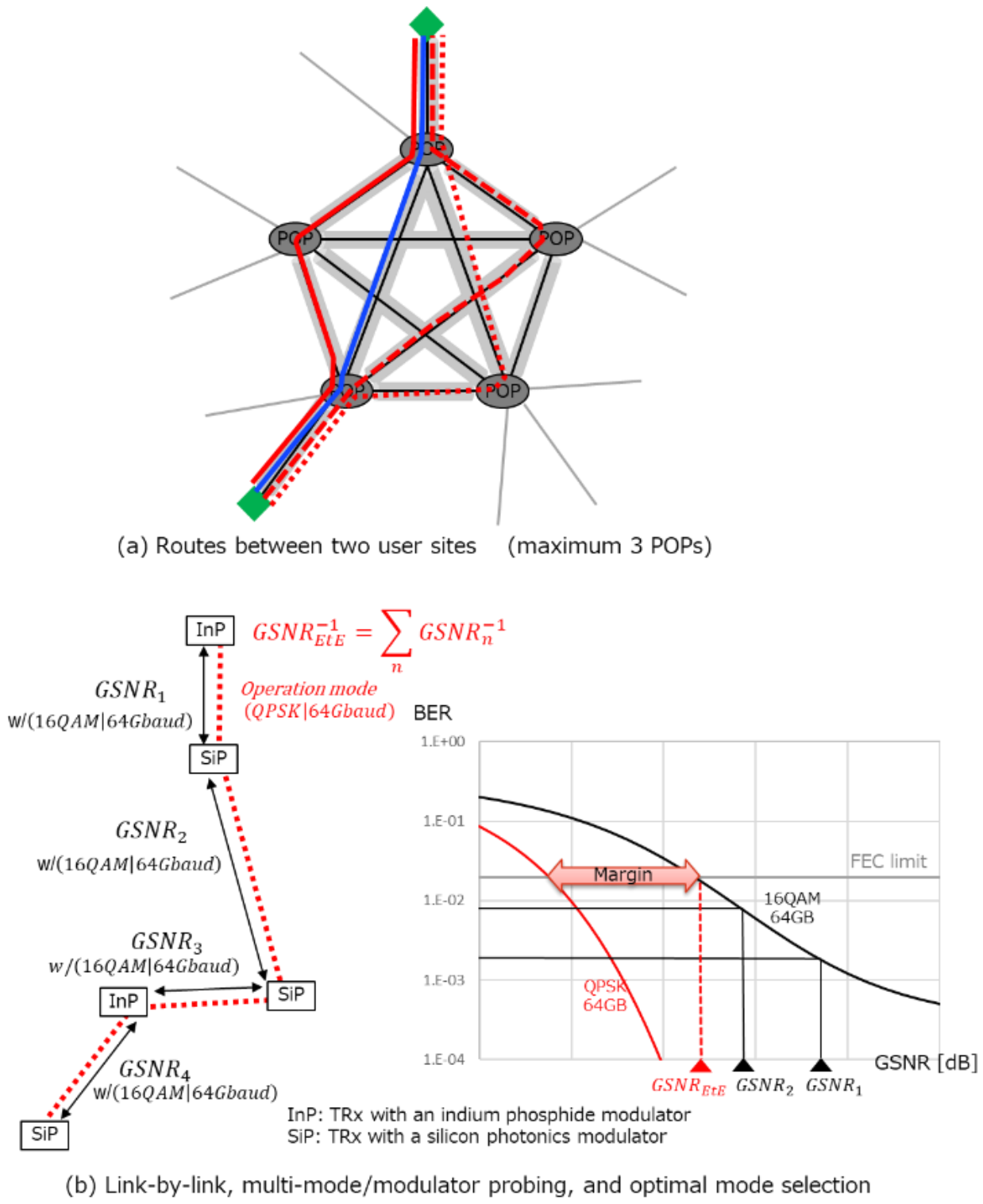

Fig. 3. Approach to selecting optimal mode with high accuracy in a short time.

Fig. 3(a) shows possible patterns in which two user sites can be connected via DCX. If the number of POPs between user sites is limited to three to minimize latency, there are four patterns as shown in Fig. 3(a) (two POPs: blue line, three POPs: three red lines). When connecting the five sites shown in Figure 2(b), there are more than 40 possible patterns. To satisfy service requirements (1)-(4) presented in Section 1, the EtE GSNR between each site must be determined accurately. The optimal mode must be selected in a short time (here, we assume that existing techniques can be applied to routing and wavelength assignment and focus our discussion in this paper on the fiber propagation design of the WDM optical path.). We have previously determined that the back-to-back characteristics of coherent TRxs can be accurately modeled following equations (1) and (2) for various modulation formats, Baud rates, and modulator types [15]. Doing so allows the Gaussian noise approximation to be applied in the metro region as well as the long haul. Equation (1) relates the BER to the back-to-back GSNR as a function of each modulation format, such as 16QAM or QPSK, and Eq. (2) shows two examples of error functions. We also previously found that the reciprocal of the GSNR is simply the sum of the reciprocal contributions of three SNR terms originating from three noise sources as equation (3): intrinsic back-to-back TRx characteristics, spontaneous emission of the amplifier, and fiber nonlinear optical effects [15]. In the previous study, we use BER and GSNR as QoT metrics. $Q^2$-factor, another widely used QoT metric, has the following one-to-one relationship with BER: $BER = \frac{1}{2}\mathrm{erfc}\left(\sqrt{\frac{Q^2}{2}}\right)$. Thus, we can use BER and $Q^2$-factor interchangeably; this paper mainly uses BER.

$$\boldsymbol{BER = \Psi_{MF}(GSNR)} \qquad \textbf{\textit{(1)}}$$

$$\Psi_{MF} = \frac{3}{8}\mathrm{erfc}\left(\sqrt{\frac{GSNR}{10}}\right) \text{ for } 16QAM, \frac{1}{2}\mathrm{erfc}\left(\sqrt{\frac{GSNR}{2}}\right) \text{ for } QPSK \qquad \textbf{\textit{(2)}}$$

$$GSNR^{-1} = SNR_{TRx}^{-1} + SNR_{ASE}^{-1} + SNR_{NLI}^{-1} \qquad \textbf{\textit{(3)}}$$

In [15], we also experimented with two types of commercial TRxs implemented with InP and SiP modulators and verified that (1)-(3) were valid for both modulators. Furthermore, we assume that the reciprocal of the EtE GSNR is equal to the reciprocal of back-to-back TRx characteristics $SNR_{TRx}$ and the sum of the reciprocal of the $GSNR_{LS,n}$ of each link, as in Eq. (4). Although it has been reported in [13] that this equation is valid with an accuracy of ±1.4 dB for 3116 km transmission as in Fig. 2(a), no experiments have yet been performed to verify the accuracy in the case of Fig. 2(b).

$$GSNR_{EtE}^{-1} = SNR_{TRx}^{-1} + \sum_n \{SNR_{ASE,n}^{-1} + SNR_{NLI,n}^{-1}\} = SNR_{TRx}^{-1} + \sum_n GSNR_{LS,n}^{-1} \qquad \textbf{\textit{(4)}}$$

Using these results, the EtE GSNR of a potential path can be estimated in a short time without directly measuring the QoT, regardless of the modulation format and modulator type of the TRx. Fig. 3(b) shows the steps to measure QoT on a link-by-link basis for one of the routes in Fig. 3(a) using a measurably convenient mode and multiple types of modulators and then calculate the EtE GSNR to select the operational mode. For example, $GSNR_1$ is measured between TRx with InP and SiP modulators. Since the QoT is relatively high for a single span, BER is measured with 16QAM 64Gbaud mode and converted to GSNR (the method for obtaining $SNR_{TRx}$ will be discussed in the last paragraph of Section 4). $GSNR_{EtE}$ is then calculated by summing the link-by-link reciprocal GSNRs, and the optimal mode can be selected after adding the operational margin [16]. In this example, $GSNR_{LS,n}$ is estimated using 16QAM 64Gbaud, and QPSK 64Gbaud is selected by adding an operational margin to the calculated $GSNR_{EtE}$. Since this approach uses common parameters (BER, modulation format, and Baud rate), it can be applied to all coherent TRx without being restricted to a specific device.

## 4. Device software architecture

The approach presented in Section 3 makes it possible to select a feasible and optimal transmission mode for fiber propagation design. However, there are still issues to be addressed in satisfying requirements (4) and (5) in Section 1 B, which will be discussed in this section. In particular, the back-to-back TRx characteristic described in Section 3 is critical for accurately estimating EtE GSNR; in the case of DCX services, the DCX operator needs to extract this characteristic from the TRx at the user site to design the optical path accurately. Later in this section, we introduce a device software architecture and procedure to improve the accuracy of GSNR and automatically optimize the mode in a short time without changing existing open interfaces and MSAs.

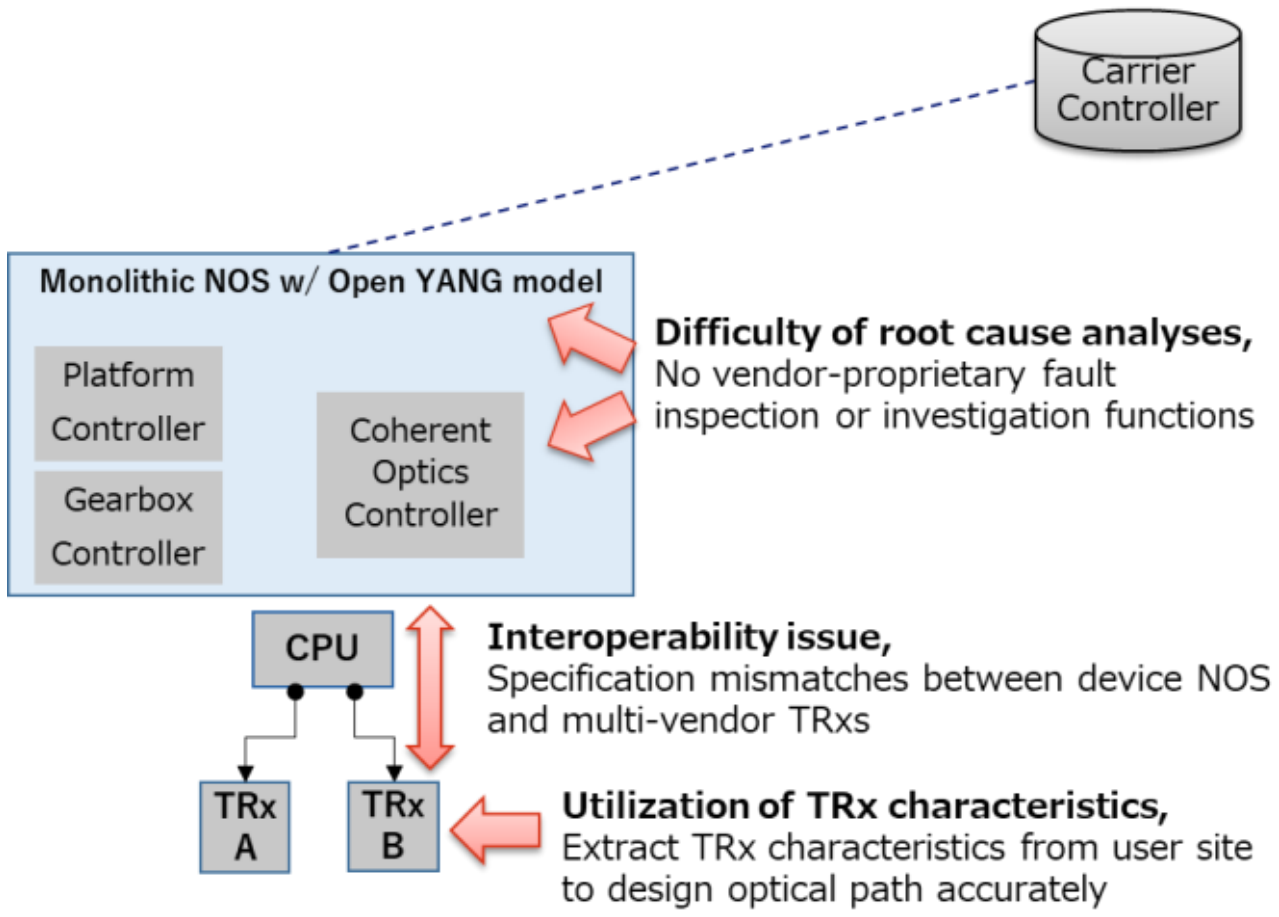

Fig. 4. Obstacles in user TRx remote control.

Figure 4 illustrates the drawbacks of when the carrier remotely controls the user TRx as a DCX servicer.

**Difficulty of root cause analyses**: If a failure occurs at a user site after the service has been launched, the carrier, as a DCX service provider, is required to isolate the failure on its own. The approach for the carrier to monitor and control the TRx at the user site would be to manage the TRx from the controller via the NOS using an open YANG model or implementing a remote monitoring and control function within the TRx. However, no open specification has been defined to monitor and control the TRx installed in user sites by utilizing vendor-proprietary fault inspection or investigation functions. Though there have been efforts to enhance traditional standardized YANG models for network design and operation automation [17], no interface has been defined for utilizing vendor-proprietary functions on these models.

**Interoperability issue**: Accidents caused by device connectivity should be prevented to ensure the safe operation of network services. For example, when using TRxs from multiple vendors, accidents can occur due to specification mismatches between the NOS and the TRx, causing the optical power input from the user device to exceed the regulated value or signals with an unexpected optical frequency being input into the network during network connection. Even for TRxs with a specified MSA, there are cases where vendors implement specifications outside the MSA to meet their customers' requirements, as well as cases where the units of parameters differ between vendors in terms of dBm or linear power (mW). Also, since each vendor TRx has a different optical frequency range, the optical frequency corresponding to f1 defined by the vendor differs for each one. For these reasons, it is difficult for carriers to remotely restore the system when a failure occurs due to a specification mismatch between the NOS and TRx at a user site.

**Utilization of TRx characteristics**: As shown by Equation (3) in Section 3, estimating the EtE GSNR requires TRx characteristics, but a new mechanism will be needed when the carrier controller designs an optical path between the TRx at the user sites.

In this paper, we propose a device software architecture to remove the obstacles shown in Figure 4 (Note that the physical network connection method for the control signal between the controller and the user muxponder and the user authentication method is outside the scope of this paper.). Figure 5 shows the device software architecture used in this experiment, which enables the carrier to monitor and control a TRx in the user muxponder. The hardware is Phoenix [5], as specified in the TIP OOPT, and the TRx is based on the Open ROADM MSA specification [3]. Two CFP2-DCOs from vendors (A and B in Fig. 5) are implemented. The IGF's disaggregated architecture described in the Open All-Photonics Network (APN) Functional Architecture [8] is used for the user muxponder and ROADM network configuration. The device software used is the NEC NOS, developed on the basis of Goldstone NOS [6] at the TIP OOPT Github site. Gearbox and coherent optics control functions are implemented as containers. In this experiment, the carrier's coherent optics controller and carrier agent (red) are added as containers in addition to the group of containers initially installed on the user muxponder (blue). The TAI architecture hides the differences between various TRx form factors or vendors. It consists of a library called libTAI which absorbs these differences and an open interface, which enables TAId to monitor and control TRx A/B agnostically in regard to form factor type or vendor type. We developed LibTAI for TRx A/B and used it in this experiment, and then performed mode optimization between various devices using the common parameters (BER, modulation format, and Baud rate). Since only common parameters are used for mode optimization, it can be done on any standard TRxs.

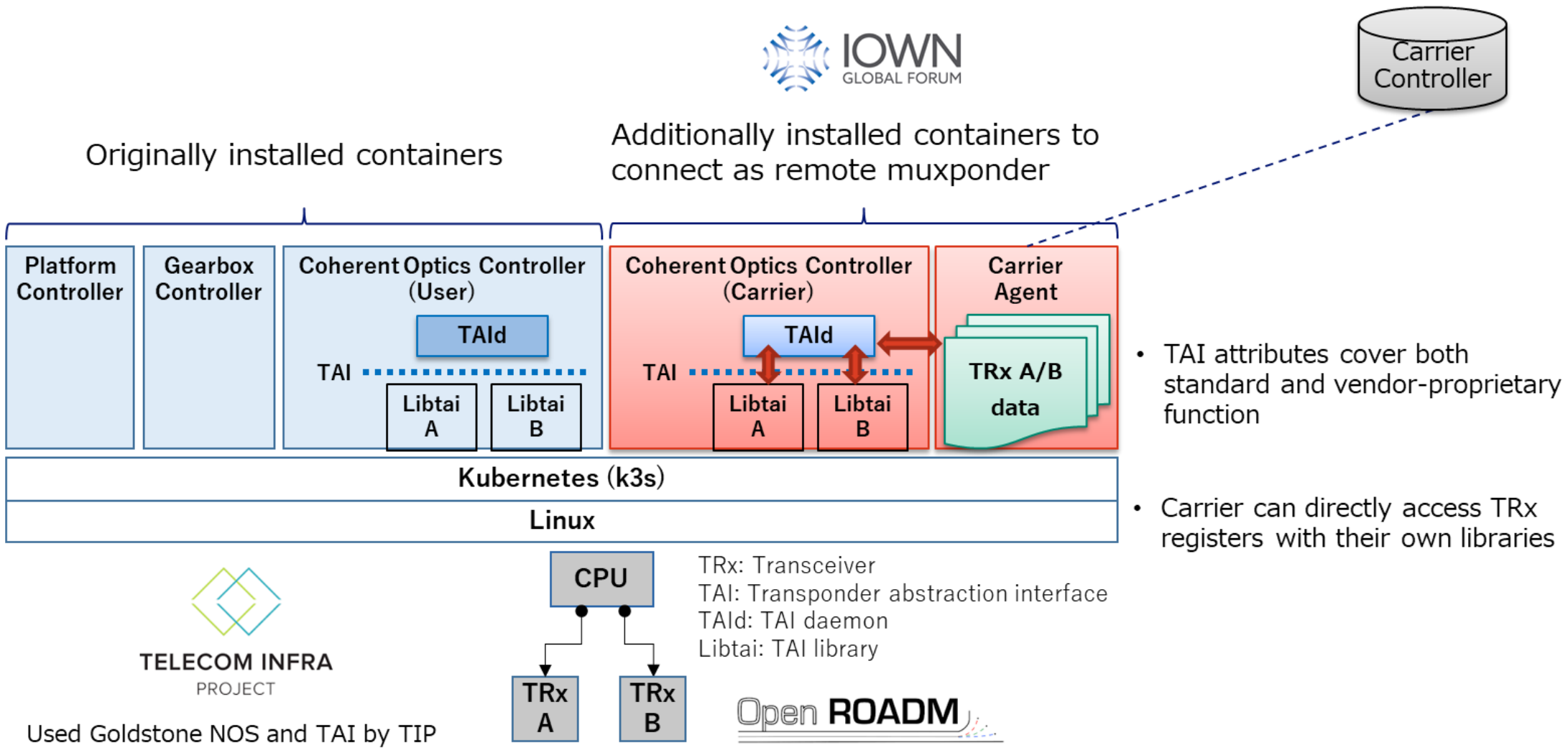

**Fig. 5:** Example of container-based device software architecture.

TAI is designed to add attributes for using vendor-proprietary functions as well as attributes for standard functions. The carrier can remotely control these containers and directly access the user's TRx registers with TAI and an in-house library to ensure the safe operation and robustness of the DCX. For example, the CFP MSA defines a module vendor private register [18]. Vendors use this to provide their proprietary functions, conducting pre-shipment inspections and root cause investigations for unexpected faults. If carrier controllers can take advantage of this capability with containers, they can use this information to conduct more advanced root cause investigations without requesting the actions of the user. Furthermore, carriers and users do not need to disclose the source code to each other when utilizing containers with executable binaries, and operational management, such as firmware updates, can be performed immediately and independently. Note that the automatic firmware upgrade of the coherent module has been demonstrated in a multi-vendor environment [19].

Fig. 6 shows an example of the user muxponder controlled by the carrier. When the user muxponder connects to the DCX, it obtains and installs the necessary container from the carrier controller and sends TRx data such as vendor name, firmware version, and parameters to the carrier controller. The carrier controller then checks to see whether the TRx is certified and requests information on the TRx characteristics from the user muxponder if it has been executed. We assume that an executed non-disclosure agreement is in place between the carrier and the TRx vendor [20]. There are several possible approaches for vendors to provide data on the TRx characteristics. For example, some TRx vendors store back-to-back BER vs. OSNR and other characteristics in registers in the TRx at the shipment. In this experiment, we stored these characteristics as a JSON file in the user muxponder. The time required to implement and complete the sequence shown in Fig. 6 was measured to be 90 seconds, which was determined to be sufficiently short.

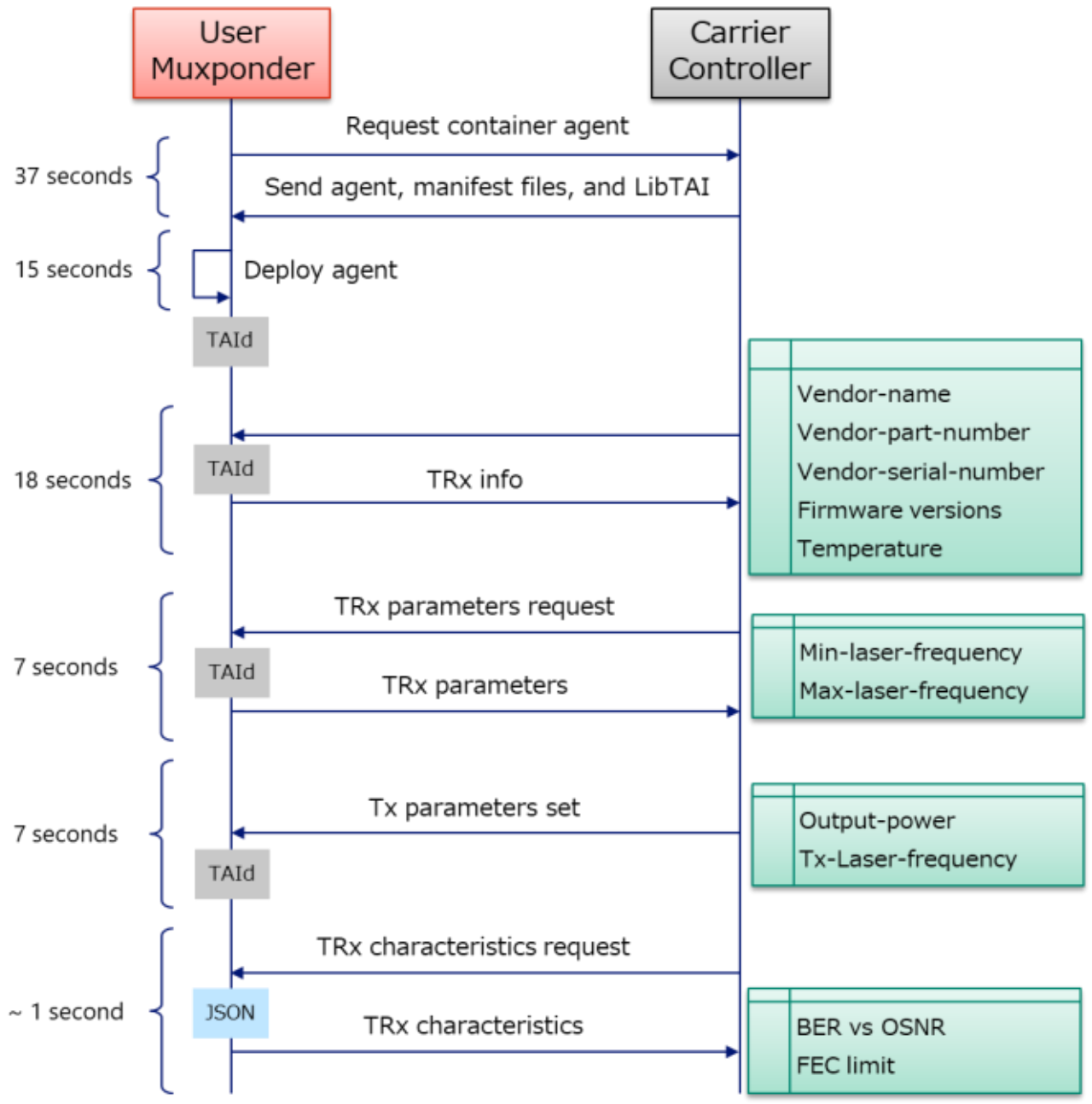

Fig. 6. Example of carrier procedure for monitoring and controlling TRx in user muxponder.

## 5. EtE GSNR measurement results and estimation errors

This section presents the evaluation results of the errors from the link-by-link measurement approach introduced in Section 3, as well as wavelength dependence and Q-value variation, using commercial open systems and field fibers.

### A. Experimental Setup

Fig. 7(a) and (b) show the field trial setup using the NSF COSMOS testbed, a city-scale programmable testbed deployed in Harlem, New York City. The COSMOS testbed provides a programmable optical networking environment, including optical space switches and Lumentum ROADM-20 units, where many experiments have been conducted [21,22,23]. We experimentally constructed two user sites, two POPs, AAL1, AAL2, and CL, with three different paths (32 km/72 km/122 km) in the C-band by utilizing commercial optical transmission devices and field fibers. We deployed white box muxponders, which comply with TIP's Phoenix requirements [5], and installed NEC's NOS, which is based on the TIP Goldstone NOS [6]. These muxponders utilize Open ROADM compliant 400G CFP2-DCO pluggable transceivers from Fujitsu Optical Components and Lumentum. Fig. 7(c) shows the Manhattan dark fiber routes we used in this experiment., i.e., 32km loop-back field fibers between Columbia University and a colocation data center at 32 Avenue of the Americas (provided by Boldyn), and an 8km loop-back field fiber between Columbia University and the City College of New York (provided by CrownCastle). The first AAL, AAL1, consists of a ROADM and a 32 km field fiber and connects to a CL. The second AAL, AAL2, consists of a ROADM and 40 km spool fiber. The CLs have different lengths and losses for different link conditions, with a total of 25 WDM background channels inserted to emulate existing in-service channels (this will be described in detail in Section C). All input powers to the fiber links were +4 dBm/channel. Two muxponders are placed at the POPs, interfacing AAL and CL for pre-FEC BER probing measurements. The controller interacts with the muxponders at the user sites and POPs through a pre-established secure channel using gRPC.

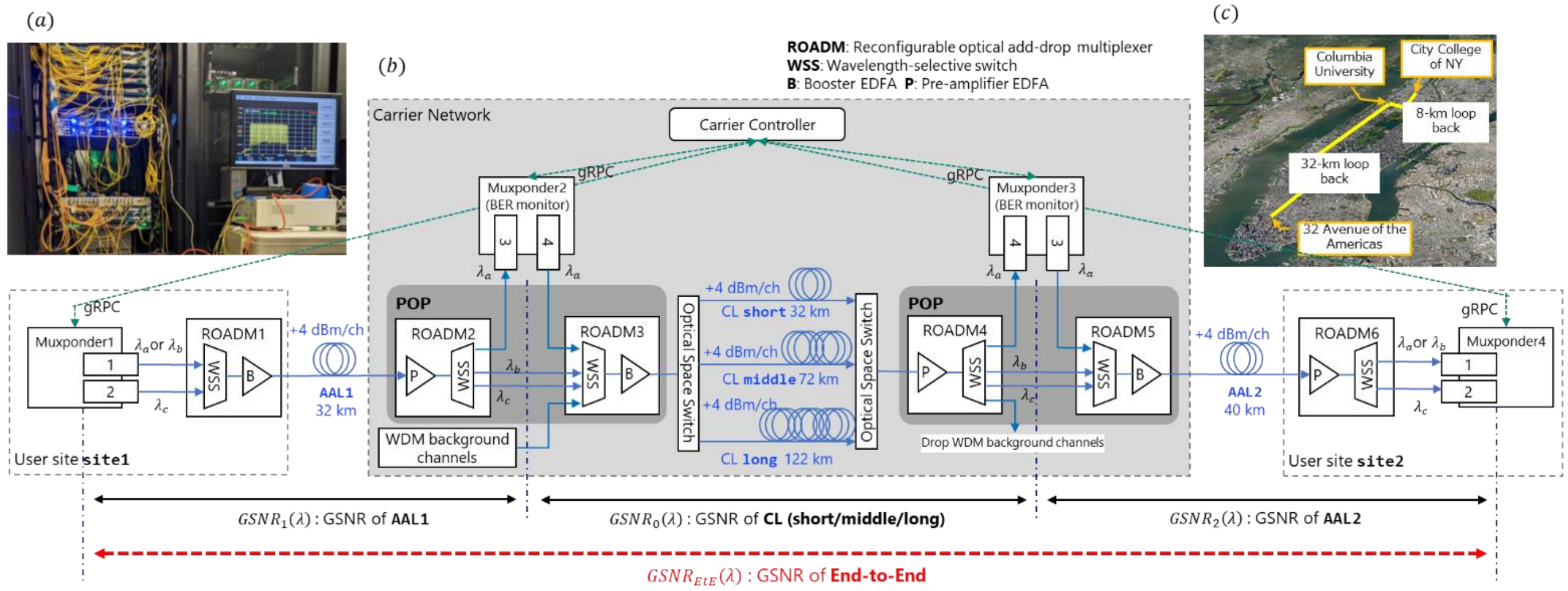

**Fig. 7:** (a) Equipment in COSMOS testbed. (b) Field trial setup. (c) Field fibers.

### B. Link-by-link Measurement Error

To determine the error of the approach shown in Fig. 3(b), we compared the measured EtE GSNR ($GSNR_{EtE}$) and estimated the GSNR with link-by-link measurement ($GSNR_1/GSNR_0/GSNR_2$) as shown in Fig. 7(b). The results are plotted in Fig. 8, where GSNR_all on the vertical axis corresponds to the entire GSNR, including the GSNR of the TRx and link.

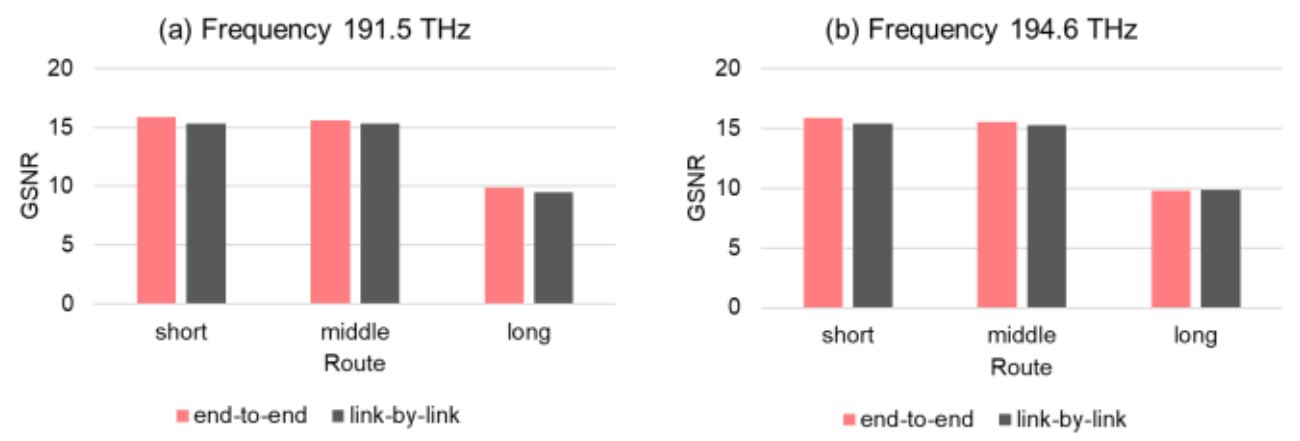

Fig. 8. Measured EtE GSNR and estimated GSNR with link-by-link measurement.

Fig. 8(a) and (b) show the difference between the two GSNRs at 191.5 THz and 194.6 THz, respectively. We changed the CL with the optical space switch and evaluated it in three route patterns: short, middle, and long. There were no significant differences in wavelength or CL distance, with errors ranging from 0.3 to 0.6 dB.

### C. Wavelength-dependent Error

Fig. 9 shows the experimental system configuration with an inset showing the WDM spectrum assignment. To measure the wavelength dependence of the link system, we established probing channels listed in the table in Fig. 9 between muxponder 2 and muxponder 3. We changed the CL with the optical space switch and evaluated it in three patterns: short, middle, and long, with 25 WDM background channels and 100 GHz spacing (see the background channels table in Fig. 9.) inserted to emulate existing in-service channels. To verify if λa (191.5 THz, at the edge of the C-band) can be utilized as a probe signal for channels of other wavelengths, we compared the difference between the GSNR measured at λa and that measured at different wavelengths (b, c, d, e). The wavelength–dependent error on GSNR probing caused by the line system was 0.05 dB, 0.11 dB, and 0.22 dB at the short, center, and long wavelengths, respectively. Although λc is located in the center of the background channels, no significant difference from λa and other wavelengths was observed. This suggests that the waveform degradation due to fiber nonlinear effects was relatively small for the system compared to the noise contribution of the TRx back-to-back characteristics, as expected for this metro-scale system.

### D. GSNR Fluctuation of System

The measured value of GSNR in the experimental system is constantly fluctuates. Since the estimation error cannot be smaller than this fluctuation, we evaluated the amount of fluctuation. We measured the GSNR every 5 seconds at three links (AAL1, AAL2, CL1) and the EtE path (EtE) and calculated the amount of fluctuation of the 5-minute moving average. We define the amount of GSNR fluctuation as the difference between the 5$^{th}$ and 95$^{th}$ percentile of the 5-minute moving average. The measurement duration was seven days.

The GSNR varied every 5 seconds by about 0.2 dB to 0.4 dB over time (Fig. 10). Since we usually use the average value over several observations, we evaluated the fluctuating amount of the moving average. We set the average window period as 5 minutes. The 5-minute moving average fluctuated from about 0.1 dB to 0.2 dB over time (Fig. 10). The amount of GSNR

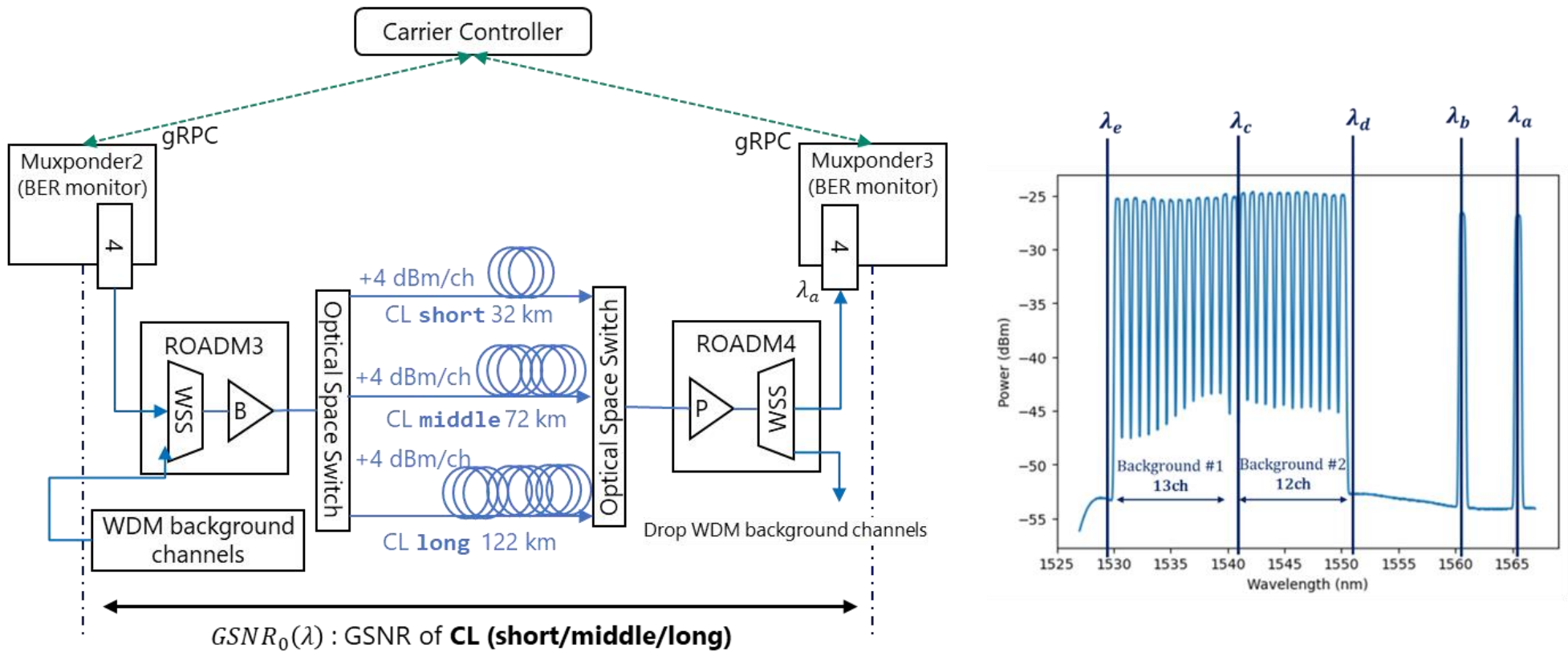

Probing channels

| name | wavelength nm | frequency THz |
|---|---|---|
| $\lambda_a$ | 1565.50 | 191.5 |
| $\lambda_b$ | 1560.61 | 192.1 |
| $\lambda_c$ | 1540.56 | 194.6 |
| $\lambda_d$ | 1550.92 | 193.3 |
| $\lambda_e$ | 1529.55 | 196.0 |

Background channels

| | start | end | No. of channels |
|---|---|---|---|
| 1st half | 193.4 | 194.5 | 12 |
| 2nd half | 194.7 | 195.9 | 13 |

Fig. 9. Experimental setup for wavelength-dependent error measurement.

fluctuations for AAL1, AAL2, CL, and EtE was 0.15, 0.10, 0.07, and 0.07 dB, respectively (Fig. 10), overall ranging from about 0.1 dB to 0.2 dB.

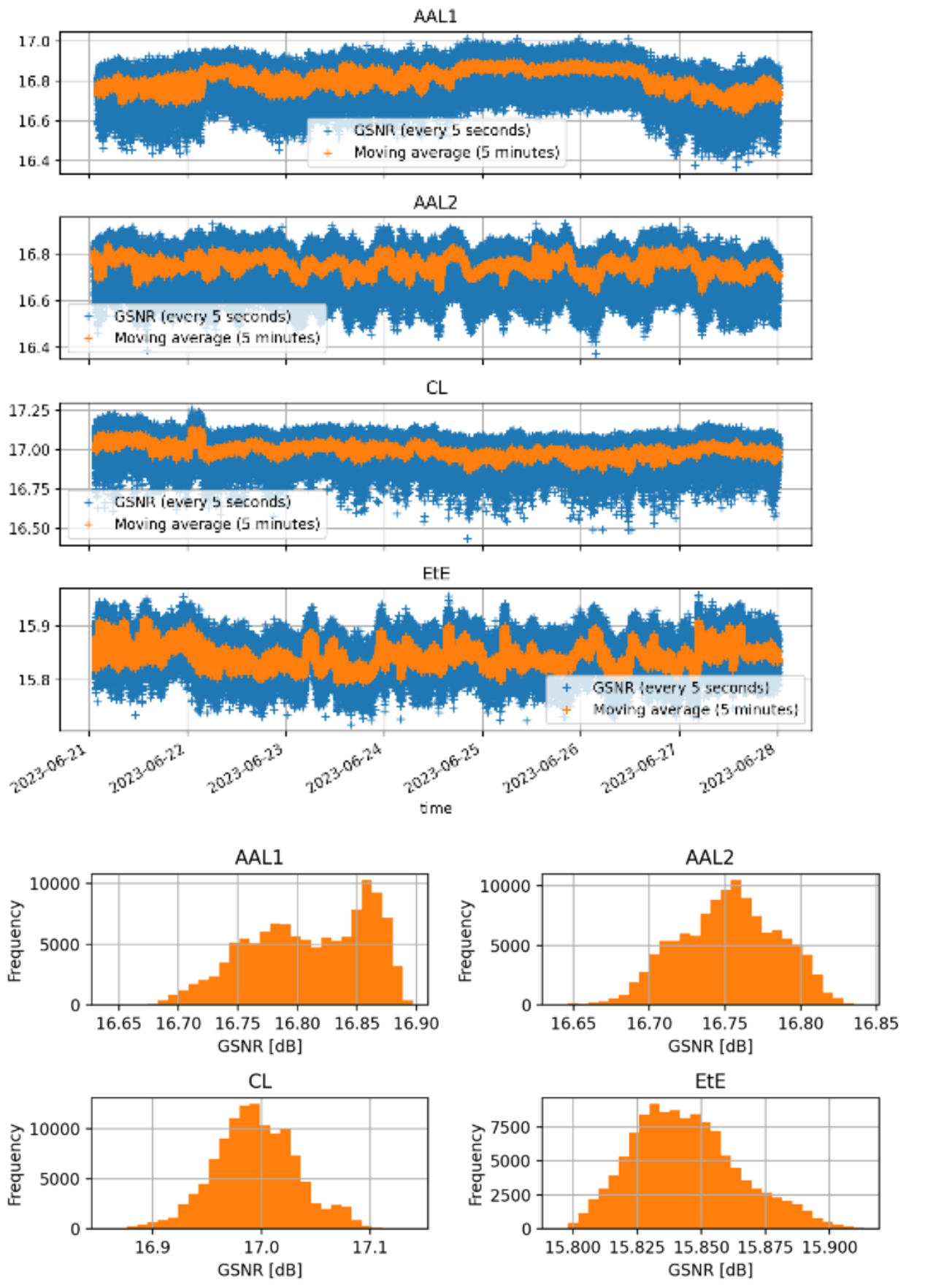

Fig. 10. Time variations of GSNR (top) and histograms of 5-minute moving average (bottom)

### E. Summary of Estimated Error

Table 1 shows the estimated error of our experimental setup in Fig. 7. The overall margin required for fast WDM provisioning using the link-by-link approach and probe light of any wavelength was 0.7 dB. This value is the same order of magnitude as the system's GSNR fluctuation, roughly 0.1 dB to 0.2 dB, so any available wavelength can be used for the approach shown in Fig. 3 with a small margin.

Tab. 1. Estimated error

| Transmission distance | LbL measurement error | WL-dependent error | Required margin |
|---|---|---|---|
| ~100 km | 0.60 dB | 0.05 dB | 0.65 dB |
| ~150 km | 0.32 dB | 0.11 dB | 0.43 dB |
| ~200 km | 0.47 dB | 0.22 dB | 0.69 dB |

## 6. Results of fast WDM provisioning experiments using field fibers

In this section, we implement the proposed architecture and evaluate the time required for estimating the quality of each link, designing the EtE optical path margin, and remotely configuring the user muxponder, for two routes (short and long) using the field fibers shown in Fig. 7(c).

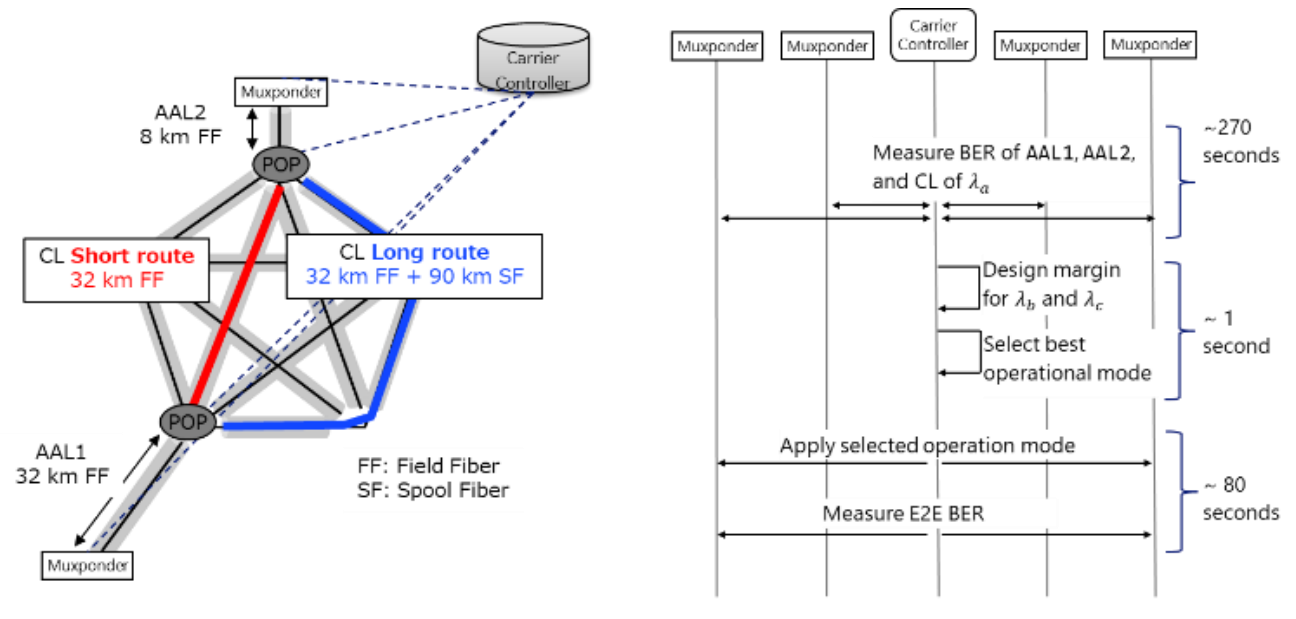

(a) Network configuration.   (b) Fast WDM provisioning overview and elapsed time.

Fig. 11. Fast WDM provisioning experiment.

Fig. 11(a) shows the experiment's network configuration. The route shown was used for AAL1, CL, and AAL2 in the experimental configuration shown in Fig. 7. Two dark fiber routes (32 km and 8 km) shown in Fig. 7(c) were used. The short CL (32 km field fiber) and long CL (32 km field fiber + 90 km spool fiber) have different lengths and losses for different link conditions, with a total of 25 WDM background channels inserted to emulate existing in-service channels (same as in Fig. 9). Fig. 11(b) shows an overview of fast WDM provisioning and the elapsed time. The controller gathers JSON-formatted TRx characteristics from all muxponders and then measures the BER of AAL1, AAL2, and CL with the muxponders. We used λa (191.5 THz) as a probing signal. The controller calculates the margins using the approach in Fig. 3(b) and selects the optimal operational mode. Finally, the controller configures the operation mode to the user muxponders and measures the EtE BER. The total provisioning time was 351 seconds, broken down as follows. Link BER measurements took 270 seconds, margin design and mode selection took under one second, and user muxponder configuration and BER measurement took 80 seconds.

We provisioned two routes, short and long, assuming a 400G bandwidth request from a user. Tab. 2 shows the measured BER and converted GSNR for AAL1, CL, and AAL2. After combining these measured GSNR values, our controller estimated the EtE GSNR and provisioned the optical paths (Tab. 3). For the short route, our controller successfully selected the single 400G DP-16QAM path using λb with a 5.2 dB secured GSNR margin. For the detour route, our controller also successfully selected the two 200G DP-QPSK paths using λb and λc as 400G DP-16QAM is not error-free due to the low EtE GSNR. The secured margins for these two paths were 5.1 dB and 4.9 dB. In all cases, the differences between the estimated and measured Q-factor (BER) was less than 0.7 dB. Most of the time is spent changing the transceiver configurations during the BER measurements. We administratively halted the transceiver for every configuration to ensure a stable transition, which required about a minute each.

Tab. 2: Measured BER (GSNR)

| Route | GSNR_ls | | |
|---|---|---|---|
| | AAL1 | CL | AAL2 |
| Short | 1.24e-3 (23.3 dB) | 1.03e-3 (23.9 dB) | 5.64e-4 (27.2 dB) |
| Long | 1.14e-3 (23.6 dB) | 1.75e-3 (12.4 dB) | 1.21e-3 (23.0dB) |

Tab. 3: Estimated GSNR EtE, selected mode, secured margin, estimated Q-factor, and measured Q-factor

| Route | Ch | GSNR_ls | Mode | Margin | Est. GSNR (Q) | Mes. GSNR(Q) | Error GSNR(Q) |
|---|---|---|---|---|---|---|---|
| Short | $\lambda_b$ | 19.7 dB | 400G | 5.2 dB | 15.47 (8.79) dB | 16.12 (9.40) dB | 0.65 (0.61) dB |
| Long | $\lambda_b$ | 11.7 dB | 200G | 5.1 dB | 10.63 (10.63) dB | 11.22 (11.22) dB | 0.61 (0.59) dB |
| Long | $\lambda_c$ | 11.6 dB | 200G | 4.9 dB | 10.63 (10.63) dB | 10.94 (10.94) dB | 0.31 (0.31) dB |

## 7. Future challenges

Although the wavelength dependence and fiber nonlinear effects were relatively small in our experimental system with the C-band, other line systems such as C+L multi-band and fibers with various characteristics will likely be routed when providing DCX services. In such cases, techniques for estimating the wavelength dependence of amplifier gain and noise figure and nonlinear effects caused by the fiber will be essential. Wavelength-dependent issues may be solved by applying the latest technology, such as transfer learning, to predict complex amplifier gain profiles [24].

DCX operators will be able to provide extended support services for their users with applying the device software architecture based on containers and TAI for streaming telemetry from TRxs to carrier controller.

We used TAI as the abstraction interface and CFP2-DCO as the coherent TRx, but the DCX needs to accommodate other form factors, such as QSFP-DD, which are widely used by data center operators. Since the source code for QSFP-DD control has been opened in SONiC [25], it should be utilized instead of TAI. An architecture that efficiently accommodates different MSAs such as CFP MSA [18], the common management interface specification (CMIS) [26], and coherent-CMIS [27] will be required.

## 8. Conclusion

We proposed a technique to quickly and accurately estimate the GSNR of all possible EtE optical paths using a QoT probe channel that measures link by link at measurable, convenient, and arbitrary modulation formats and modulators. We also proposed an approach to select the optimal mode in a short time based on Gaussian noise approximation with a device software architecture that utilizes server-based technologies such as containers and the hardware abstraction interface for coherent modules. Since this approach uses common parameters (BER, modulation format, and Baud rate), it can be applied to all coherent TRxs without being restricted to a specific device. These techniques enable DCX services that directly connect many-to-many user sites deployed in a two-dimensional plane via optical paths to respond to sudden traffic demands and quickly reconstruct optical path routes in the event of a disaster. We identified issues related to user support/responsibility and interoperability/network robustness and proposed a Linux-based muxponder device software architecture with a DCX operator agent as a container. We experimentally constructed three different paths by utilizing commercial optical transmission devices and field fibers; the accuracy estimated from the sum of the measurements for each link was 0.6 dB, and the wavelength-dependent error was about 0.3 dB. We combined WDM provisioning with the device software architecture in which users and carriers cooperate to optimize optical path transmission modes. Finally, we evaluated the time required to estimate the link quality for designing the EtE optical path margin and remotely configuring user muxponders. We provisioned two routes utilizing field fibers at 32 km and 122 km distances, assuming a 400G bandwidth request from a user. Our controller successfully selected the single lambda 400G DP-16QAM path for the short route, and the two lambda 200G DP-QPSK paths for the long route within 6 minutes automatically. The estimated and measured Q-factor (BER) differences were less than 0.7 dB.

**Funding Information.** NSF grants CNS-1827923, OAC-2029295, CNS-2112562, CNS-2128638, EEC-2133516, CNS-2211944, NSF Grant CNS-2148128 and by funds from federal agency and industry partners as specified in the Resilient & Intelligent NextG Systems (RINGS) program, and Science Foundation Ireland under Grant #13/RC/2077 P2.

**Acknowledgment**. We thank Rob Lane and the CRF team (Columbia), Lumentum, FOC, and the Telecom Infra Project OOPT-PSE working group.

## References

1. IOWN Global Forum. (2023). Open APN architecture PoC reference, [Online]. Available: https://iowngf.org/technology/ (visited on 08/05/2023).
2. Optical Internetworking Forum. (2023). 400ZR, Management/CMIS, [Online]. Available: https://www.oiforum.com/ (visited on 09/06/2023).
3. OpenROADM MSA. (Aug. 2022). Open ROADM MSA specification ver 5.1, [Online]. Available: https : / / openroadm.org/ (visited on 5/01/2023).
4. OpenZR+. (2023). [Online]. Available: https://openzrplus.org/ (visited on 09/06/2023).
5. Telecom Infra Project. (2023). Phoenix: An open white-box L0/L1 transponder, [Online]. Available: https://telecominfraproject.com/oopt/ (visited on 08/05/2023).
6. Telecom Infra Project. (2023). Goldstone: Goldstone open NOS, [Online]. Available: https://github.com/oopt-goldstone/goldstone-mgmt (visited on 05/01/2023).
7. Telecom Infra Project. (2023). TAI: Transponder abstraction interface, [Online]. Available: https : / / github.com/Telecominfraproject/oopt-tai (visited on 04/27/2023).
8. IOWN Global Forum. (2023). Open All-Photonic Network Functional Architecture, [Online]. Available: https://iowngf.org/technology/ (visited on 08/05/2023).
9. P. Samadi, M. Fiorani, Y. Shen, et. al., "Flexible architecture and autonomous control plane for metro-scale geographically distributed data centers," J. Lightwave Technology, vol. 35, no. 6, March 15, 2017.
10. M. Ju, Y. Liu, F. Zhou, et al., "Disaster-resilient and distance-adaptive services provisioning in elastic optical inter-data center networks," J. Lightwave Technology, vol. 40, no. 13, July 1, 2022.
11. K. Kaeval, T. Fehenberger, J. Zou, et al., "QoT assessment of the optical spectrum as a service in disaggregated network scenarios," J. Opt. Commun. Netw. Vol. 13, no. 10, Oct. 2021.
12. K. Kaeval, S. L. Jansen, F. Spinty, et al., "Characterization of the optical spectrum as a service," J. Opt. Commun. Netw. Vol. 14, no. 5, pp. 398-410, 2022.
13. K. Kaeval, J. Myyry, K. Grobe, et al., "Concatenated GSNR profiles for end-to-end performance estimations in disaggregated networks," in Optical Fiber Communication Conference (2022).
14. H. Nishizawa, T. Sasai, T. Inoue, et al., "Dynamic optical path provisioning for alien access links: Architecture, demonstration, and challenges," IEEE Communications Magazine, pp. 1–7, 2023. DOI: 10 .1109/MCOM.006.2200567.
15. T. Mano, A. D'Amico, E. Virgillito, et al., "Modeling transceiver ber-osnr characteristic for qot estimation in short-reach systems," in Proceedings of the International Conference on Optical Network Design and Modeling (ONDM), to appear, 2023.
16. V. Curri, "GNPy model of the physical layer for open and disaggregated optical networking [Invited]," JOCN, Vol. 14, Issue 6, pp. C92-C104 (2022).
17. Openconfig. (2023). Terminal device properties found issues – New release proposed v0.2.0, [Online] Available: https://github.com/openconfig/public/issues/910 (visited on 08/30/2023).

18. CFP MSA Management Interface Specification Version 2.6, March 24, 2017, https://cfp-msa.org/wp-content/uploads/2022/10/CFP_MSA_MIS_V2p6r06a.pdf
19. C. Qin, B. Guan, K. Edwards, et al,, "Interoperable 400ZR Deployment at Cloud Scale," in Optical Fiber Communication Conference (OFC) 2023, Optica Publishing Group, 2023, W3H.2.
20. A. Ferrari, M. Filer, K. Balasubramanian, et al., "GNPy: an open source application for physical layer aware open optical networks," vol. 12, issue 6, pp. C31-C40 2020.
21. D. Raychaudhuri, I. Seskar, G. Zussman, et al., "Challenge: COSMOS: A city-scale programmable testbed for experimentation with advanced wireless," in Proc. ACM MobiCom'20, 2020
22. T. Chen, J. Yu, A. Minakhmetov, et al., "A software-defined programmable testbed for beyond 5g optical-wireless experimentation at city-scale", IEEE Network, vol. 36, no. 2, pp. 90–99, 2022. DOI: 10. 1109 / MNET. 006. 2100605.
23. Y. K. Huang, Z. Wang, E. Ip, et al., "Field trial of coexistence and simultaneous switching of real-time fiber sensing and 400GbE supporting DCI and 5G mobile services," in Proc. IEEE/OPTICA Optical Fiber Communication Conference (OFC'23), W3H.4, Mar. 2023.
24. Z. Wang, D. Kilper, and T. Chen, "Transfer learning based roadm edfa wavelength-dependent gain prediction using minimized data collection," in Optical Fiber Communication Conference (OFC) 2023, Optica Publishing Group, 2023, Th2A.1.
25. Linux Foundation project. (2023). SONiC: Software for Open Networking in the Cloud, [Online]. Available: https://sonicfoundation.dev/ (visited on 09/1/2023).
26. Common Management Interface Specification (CMIS) Revision 5.2, April 27, 2022, https://www.oiforum.com/wp-content/uploads/OIF-CMIS-05.2.pdf
27. Implementation Agreement for Coherent CMIS, OIF-C-CMIS-01.0, Jan. 14, 2020, https://www.oiforum.com/wp-content/uploads/OIF-C-CMIS-01.0.pdf